\documentclass[12pt]{article}
\usepackage{amsmath}
\usepackage{graphicx}
\date{}

\def\p{Painlev\'e}

\newcommand{\beq}{\begin{equation}}
\newcommand{\eeq}{\end{equation}}
 
\medskip

\def\d{\delta}

\def\z{\zeta}
\def\x{\xi}

\def\o{\omega}

\def\l{\lambda}
\def\m{\mu}
\def\p{\pi}
\def\r{\rho}
\def\i0{\int_0^{\infty}}

\def\p1{\frac{\partial}{\partial x_1}}
\def\p2{\frac{\partial}{\partial x_2}}
\def\p3{\frac{\partial}{\partial x_3}}

\def\d3{\frac{d}{d x_3}}
\def\q1{\frac{\partial}{\partial x_1}}
\def\q2{\frac{\partial}{\partial x_2}}
\def\z1{\frac{\partial}{\partial x_1}}

\begin{document}

\title{ Riemann-Hilbert Approach to the Elastodynamic Equation. Half plane}

\author{Alexander Its, Elizabeth Its \\
{\it Department of Mathematical Sciences},\\ 
{\it Indiana University -- Purdue University  Indianapolis}\\
{\it Indianapolis, IN 46202-3216, USA}}

\maketitle


\begin{abstract}

We show, how the Riemann-Hilbert approach to the elastodynamic 
equations, which have been suggested in our preceding papers, 
works in the half-plane case. We pay a special attention to
the appearance of the Rayleigh waves within  the scheme.

\end{abstract}

\section{Introduction}

\vskip 0.3in

This paper is a complement to our previous  work \cite{iik} where,
following the general ideas of Fokas' method \cite{g3} -\cite{g55},
we started to develop the Riemann-Hilbert scheme for solving 
the elastodynamic equations in the wedge-type domains. Our
first goal is to analyze the quarter plane case. In \cite{iik}, we
show that the problem  can be reduced to the solution of a
certain matrix, $2\times 2$ Riemann-Hilbert problem with a shift
posed on a torus. A detail analysis of this problem is our ultimate
goal which we hope to be able to present  in our further  publications.
The aim of this paper is much more modest. We want to show how
the basic ingredients of the elasticity  theory, such as the Rayleigh waves, 
are produced in the framework  of Fokas' method. To this end 
we shall consider the simplest case, the problem in the half-plane.
Of course, the problem can be easily solved via the standard 
separation of variables. However, its analysis in the framework
of the Riemann-Hilbert method shows many of the features which
are also present in the more interesting and important case
of the quarter plane.

The quarter plane case has already been outlined in \cite{iik}.
In the next section  we shall remind what we did there.

\section{Lax Pair for the elastodynamic equation}\label{sec1}

The elastodynamic equation
in an isotropic medium defined by the Lam\'e parameters 
 $ \l, \m $,  density $\r$ and frequency $\o$ can be written as the
 following system of two scalar equations.
 \beq\label{sc01}
u_{xx}+\frac{h^2}{l^2}u_{zz}+\frac{l^2-h^2}{l^2}w_{xz}+h^2u=0
\eeq
\beq\label{sc011}
w_{zz}+\frac{h^2}{l^2}w_{xx}+\frac{l^2-h^2}{l^2}u_{xz}+h^2w=0
\eeq
where   $h^2=\frac{\rho \o ^2}{ \l+2\m},$  $l^2=\frac{\rho \o
^2}{\m}$. The problem is two - dimensional in $xz$ plane, and $u$ and $w$
are the $x$ and $z$ components of displacement, respectively.  
For the half plane  problem ($z\geq 0$) on the  surfaces $z=0$ the stress free
boundary  conditions are :
\beq\label{sc4}
T_{xz}=\m \left(  u_z + w_x\right) =-T^{(0)}_{xz},\quad T_{zz}=\l u_x+
(\l+2\m) w_z=-T^{(0)}_{zz},\quad z=0
\eeq
where $T^{(0)}_{zz}$ and $T^{(0)}_{xz}$  denote 
the given stresses which could be interpreted for example as the stresses of the incident Rayleigh wave. 
The solution should also satisfy Sommerfeld's radiation conditions \cite{g6} which we shall specify
latter on (see equation (\ref{som}) below).

In \cite{iik},  following the methodology of  \cite{g44}, we showed that equations (\ref{sc01}), (\ref{sc011})
are compatibility conditions of the following two {\it Lax pairs}, written for the auxiliary  scalar functions,
$\phi \equiv \phi(z,x;k)$ and  $\psi \equiv \psi(z,x;k)$ (see \cite{iik} and \cite{g57} for details),
\beq\label{lp4022}
 \phi _z -ik\phi = \frac{1}{h^2}(\sqrt{k^2-h^2}-k)\tau _1-\frac{1}{h^2}
(\tau_{1x}+i\tau_{1z}),
\eeq
$$ \phi _x +\sqrt{k^2-h^2} \phi =\frac{i}{h^2}(k-\sqrt{k^2-h^2})\tau _1-\frac{i}{h^2}
(\tau_{1x}+i\tau_{1z}),
$$
and
\beq\label{lp402233}
 \psi _z -ik\phi = \frac{1}{l^2}(\sqrt{k^2-l^2}-k)\tau _2-\frac{1}{l^2}
(\tau_{2x}+i\tau_{2z}),
\eeq
$$ \psi _x +\sqrt{k^2-l^2} \phi =\frac{i}{l^2}(k-\sqrt{k^2-l^2})\tau _2-\frac{i}{l^2}
(\tau_{2x}+i\tau_{2z}),
$$
where $\tau_1(z,x)$ and $\tau_2(z,x)$ are the lame potentials given by the equations,
\beq\label{potential1}
 \tau_1 = \frac{1}{2}(u_x + w_z),\quad \tau_2 = \frac{1}{2}(w_x - u_z).
\eeq
and $\phi$  and $\psi$  satisfy the following asymptotic conditions ,
\beq\label{lp4024}
\phi, \psi = O\left(\frac{1}{k}\right), \quad k\rightarrow \infty^+
\eeq
$$\phi, \psi = O(1),  \quad k\rightarrow \infty^-.$$
The last condition in conjunction with systems (\ref{lp4022}), (\ref{lp402233}) yields in fact
the more specific asymptotic representation  of the solutions $\phi$  and $\psi$ as $
k \rightarrow \infty^-$. Indeed we have (cf. \cite{g57}), that
\beq\label{cond1}
\phi = -\frac{2i}{h^2}\tau_1 + O\left(\frac{1}{k}\right), \quad \quad 
\psi = -\frac{2i}{l^2}\tau_2 + O\left(\frac{1}{k}\right) \quad k\rightarrow \infty^-.
\eeq
In these formulae, $k\rightarrow \infty^{\pm}$ means that $k\rightarrow \infty$ and $\sqrt{k^2 -h^2}, \sqrt{k^2 -l^2}
\rightarrow \pm k + É.$.

Introducing  the new spectral parameter $\zeta$
 as follows, 
\beq\label{uif41} 
k = \frac{h}{2}\left(\zeta +\frac{1}{\zeta}\right),\quad  \sqrt{k^2-h^2} = \frac{h}{2}\left(\zeta -\frac{1}{\zeta}\right),
\eeq
so that,
$$
\;\zeta \rightarrow \infty \;{\mbox{as}}\;
k \rightarrow \infty ^+ \;{\mbox{and }}\; \zeta \rightarrow 0 \;{\mbox{as}}\;
k \rightarrow \infty ^- \,,  
$$
one can rewrite the first Lax pair (\ref{lp4022}) as
\beq\label{uif411}
\phi_{z}-\frac{ih}{2}\left(\zeta +\frac{1}{\zeta}\right)\phi=Q_1
\eeq
\beq\label{uif412}
\phi_{x}+\frac{h}{2}\left(\zeta -\frac{1}{\zeta}\right)\phi={\tilde Q}_1
\eeq
where $Q_1,\;{\tilde Q}_1$ are the right-hand side parts of (\ref{lp4022}). In
terms of $\zeta$ they are :
\beq\label{uif413}
Q_1=-\frac{\tau_1}{\zeta h}-\frac{1}{h^2}(\tau_{1x}+\tau_{1z});\quad
{\tilde Q}_1=\frac{i \tau_1}{\zeta h}-\frac {i}{h^2}(\tau _{1x}+i\tau _{1z}).
\eeq
The normalization conditions (\ref{lp4024}) and (\ref{cond1}) in terms of
the new variable $\zeta$ read,
\beq\label{cond1new}
\phi = O\left(\frac{1}{\zeta}\right), \quad \zeta\rightarrow \infty,
\eeq
\beq\label{cond2}
\phi = -\frac{2i}{h^2}\tau_1 + O(\zeta) \quad\zeta\rightarrow 0.
\eeq

The new spectral parameter for the second Lax pair we shall denote
$\tilde{\zeta}$. The variable $\tilde{\zeta}$ is defined by the relations,
\beq\label{zetatilde} 
k = \frac{l}{2}\left(\tilde\zeta +\frac{1}{\tilde\zeta}\right),\quad  
\sqrt{k^2-l^2} = \frac{l}{2}\left(\tilde\zeta -\frac{1}{\tilde\zeta}\right),
\eeq
so that,
$$
\;\tilde\zeta \rightarrow \infty \;{\mbox{as}}\;
k \rightarrow \infty ^+ \;{\mbox{and }}\; \tilde\zeta \rightarrow 0 \;{\mbox{as}}\;
k \rightarrow \infty ^- \,. 
$$
The second Lax pair reads as follows
\beq\label{uif4111}
\psi_{z}-\frac{il}{2}\left(\tilde\zeta +\frac{1}{\tilde\zeta}\right)\psi=Q_2,
\eeq
\beq\label{uif4121}
\psi_{x}+\frac{l}{2}\left(\tilde\zeta -\frac{1}{\tilde\zeta}\right)\psi={\tilde Q}_2,
\eeq
where
\beq\label{Q2}
Q_2=-\frac{\tau_2}{\tilde\zeta l}-\frac{1}{l^2}(\tau_{2x}+\tau_{2z});\quad
{\tilde Q}_2=\frac{i \tau_2}{\tilde\zeta l}-\frac {i}{l^2}(\tau _{2x}+i\tau _{2z}),
\eeq
and it is supplemented by the normaliztion conditions,
\beq\label{cond1new2}
\phi = O\left(\frac{1}{\tilde\zeta}\right), \quad \tilde\zeta\rightarrow \infty,
\eeq
\beq\label{cond3}
\phi = -\frac{2i}{l^2}\tau_2 + O(\tilde\zeta) \quad\tilde\zeta\rightarrow 0.
\eeq

The potentials  $\tau_1$ and $\tau_2$ can be taken as the basic objects instead of
the original displacements $u$ and $w$. Indeed, as it follows  from
(\ref{sc01}) and (\ref{sc011}), the functions $u$ and $w$ can be reconstructed 
via $\tau_1$ and $\tau_2$ with the help of the following equations, 
\beq\label{tauuw}
u = -\frac{2}{h^2}\tau_{1x} +\frac{2}{l^2}\tau_{2z},\quad\mbox{and}\quad
w = -\frac{2}{h^2}\tau_{1z} -\frac{2}{l^2}\tau_{2x},
\eeq
respectively. Also, in terms of potentials $\tau_1$ and $\tau_2$,  Sommerfeld's radiation conditions 
can be written as
\beq\label{som}
\lim_{R\rightarrow \infty}R\left(\frac{\partial\tau_1}{\partial R} -ih\tau_1\right) = 0,
\quad\lim_{R\rightarrow \infty}R\left(\frac{\partial\tau_2}{\partial R} -il\tau_1\right) = 0,
\quad R = \sqrt{x^2+z^2}.
 \eeq

The potentials $\tau_{1,2}$ satisfy the Helmholtz equations,
\beq\label{Helmholtz1}
\tau_{1xx} + \tau_{1zz} +h^2\tau_1 = 0, 
\eeq
\beq\label{Helmholtz2}
\tau_{2xx}+\tau_{2zz} +l^2\tau_1 = 0.
\eeq
This fact  follows again from the basic elastodynamic system (\ref{sc01}) - (\ref{sc011}).
Hence the linear  systems  (\ref{uif411})-(\ref{uif412}) and (\ref{uif4111})-(\ref{uif4121})
can be thought of as the Lax pairs for the  equations (\ref{Helmholtz1})  and 
(\ref{Helmholtz2}), respectively. This 
Lax pair representation of the Helmholtz equation has already been known
and used for the analysis of the boundary value problem for the Helmholtz
equation in \cite{g5} and \cite{g57}.   The very important novelty of the situation we are dealing with
in this paper is that the boundary conditions, which relations (\ref{sc4}) and (\ref{sc5})
impose on the functions $\tau_1$ and $\tau_2$  are completely different from 
the ones which appear in the pure Hemholtz problem. The most distinct feature
of these conditions is that they mix the two Lax pairs together, and this in turn
complicates dramatically the analysis of the global relation (the main ingredient 
of Fokas' method \cite{g55})  in the case of the quarter space. In the half space,
however, the solution of the global relation can be obtained in  the closed
form and by simple algebraic means.

\section{Half space problem}\label{sec2}

The considerations of the previous section were general. We now
apply the Lax pair representation of the elastodynamic equation to the
half plane problem. We will basically repeat the constructions of the 
Section 3 of \cite{iik}.

\subsection{Integration of the  Lax Pairs. The integral representation for the potential functions.}

Rewriting (\ref{uif411}, \ref{uif412}) as,
\beq\label{uif0411}
e^{\frac{ih}{2}(\zeta +\frac{1}{\zeta})z-\frac{h}{2}(\zeta -\frac{1}{\zeta})x}
(\phi e^{-\frac{ih}{2}(\zeta +\frac{1}{\zeta})z+\frac{h}{2}(\zeta
-\frac{1}{\zeta})x})_z =Q_1
\eeq
\beq\label{uif1411}
e^{\frac{ih}{2}(\zeta +\frac{1}{\zeta})z-\frac{h}{2}(\zeta -\frac{1}{\zeta})x}
(\phi e^{-\frac{ih}{2}(\zeta +\frac{1}{\zeta})z+\frac{h}{2}(\zeta
-\frac{1}{\zeta})x})_x ={\tilde Q}_1.
\eeq
and  integrating yields the following general formula for the solution of  (\ref{uif411}, \ref{uif412}).
\beq\label{uif2411}
\phi(\zeta,x,z)=e^{\frac{ih}{2}(\zeta +\frac{1}{\zeta})z-\frac{h}{2}(\zeta
-\frac{1}{\zeta})x}\int_{(x^*,z^*)}^{(x,z)}
 e^{-\frac{ih}{2}(\zeta +\frac{1}{\zeta})z'+\frac{h}{2}(\zeta
-\frac{1}{\zeta})x'}[Q_1dz'+{\tilde Q}_1dx']
\eeq
It is worth noticing, that the  path independence  of the line integral in the right hand side
is equivalent to the elastodynamic equation. 

\begin{figure}[ht]
\begin{center}
\resizebox{3in}{!}{ 
\includegraphics{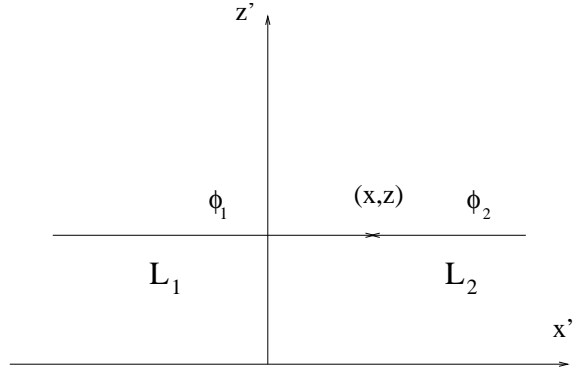}
}
\caption{Contours $L_i$ and the solutions $\phi _i(x,z) ,\; i=1,2$
of the first scalar Lax pair}
\label{f1}
\end{center}
\end{figure}
Choosing the contours of integration as shown
in Figure \ref {f1} one obtains two distinct  solutions:

\beq\label{uif2400}
\phi _1(\zeta,x,z) = \int_{-\infty }^x e^{\frac{h}{2}(\zeta -\frac{1}{\zeta})(x'-x)} {\tilde
Q}_1(\zeta,z,x') dx'
\eeq
 \beq\label{phi2}
\phi _2(\zeta,x,z) = \int_\infty^x e^{
\frac{h}{2}(\zeta -\frac{1}{\zeta})(x'-x)} {\tilde Q}_1(\zeta,z,x')
dx'  
\eeq
\begin{figure}[ht]
\begin{center}
\resizebox{3in}{!}{ 
\includegraphics{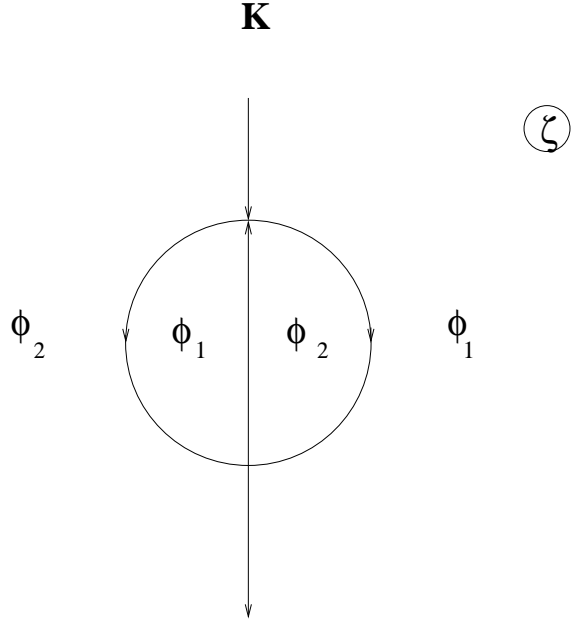}
}
\caption{Regions of analyticity of functions $\phi _i,\; i=1,3$
of the first scalar Lax pair}
\label{f2}
\end{center}
\end{figure}
The functions $\phi _1, \,  \phi _2$  are analytic in the regions
of the complex $\zeta$ plane which are shown in Figure \ref{f2}. 
The difference, 
$$
\phi _1 -\phi _2,
$$
is the solution of the homogeneous version of system  (\ref{uif411}, \ref{uif412}).
Therefore, 
\beq\label{jump1}
\phi _1 -\phi _2= e^{ih(\zeta +\frac{1}{\zeta})z/2-h(\zeta -\frac{1}{\zeta})x/2}\rho
_{12},
\eeq
where the jump function $\rho _{12}(\zeta),\; j,k= 1,2$  does not depend on $x$ and $z$ and,
as a function of $\zeta$, is well defined on
the boundaries of the regions in Figure \ref{f2}, that is on the oriented contour $K$ also depicted
in Figure \ref{f2}. 

A key point now is to look at  relation (\ref{jump1}) as at the {\it Riemann-Hilbert problem}
of finding  the piecewise analytic function $\phi(\zeta)$ whose boundary values on the
contour $K$, i.e.
$ \phi_{+} = \phi_1, \quad \phi_{-} = \phi_2$,
satisfy the jump relation (\ref{jump1}). Solving this Riemann-Hilbert
problem  we obtain the following integral representation
for the piece-wise analytic   function $\phi_\zeta)$,
\beq\label{uif42}
\phi(\zeta)= \frac{1}{2\pi i}\int_K \frac{e^{\frac{ih}{2}(s+1/s)z - 
\frac{h}{2}(s-1/s)x}}{s-\zeta} \rho_{12}(s)ds.
\eeq
Taking into account (\ref{cond2}), we derive from (\ref{uif42})  the integral representation
for $\tau _1$,
\beq\label{uif43}
\tau _1\equiv \frac{1}{2} (u_x+w_z)=\frac{h^2}{4\pi}\int_K
\frac{e^{\frac{ih}{2}(\zeta+1/\zeta)z - 
\frac{h}{2}(\zeta-1/\zeta)x}}{\zeta} \rho_{12}(\zeta)d\zeta.
\eeq
Similar representation we obtain for the potential $\tau_2$ using the second Lax pair,
\beq\label{uif4333}
\tau _2\equiv \frac{1}{2} (w_x-u_z)=\frac{l^2}{4\pi}\int_K
\frac{e^{\frac{il}{2}(\tilde\zeta+1/\tilde\zeta)z - 
\frac{h}{2}(\tilde\zeta-1/\tilde\zeta)x}}{\tilde\zeta} \tilde\rho_{12}(\tilde\zeta)d\tilde\zeta.
\eeq

To complete the  solution of the half space problem, we only need now to express 
the jump function $\rho_{12}(\zeta)$ and the similar function, $\tilde{\rho}_{12}(\tilde\zeta)$
(coming from the second Lax pair),
in terms of the given boundary data, i.e. in terms of the stresses  $T_{xz}^{(0)}$ and  $T_{zz}^{(0)}$.
To this end, we notice  that
equation (\ref{jump1}) holds for all $x$ and $z$ and that $\rho_{12}(\zeta)$ does not
depend on $x$ and $z$; therefore, using this equation
for $x=0$ and $z=0$ and remembering  the definitions (\ref{uif2400}), (\ref{phi2})
of the solutions $\phi_{1,2}$,  one obtains the following formula for the jump function $\rho_{12}$,
\beq\label{uif40333}
\rho_{12}(\zeta)= \int_{-\infty}^\infty e^{
\frac{h}{2}(\zeta -\frac{1}{\zeta})x'} {\tilde Q}_1(\zeta,0,x')dx'.
\eeq 
The integrand, ${\tilde Q}_1(\zeta,0,x')$ involves the boundary values  of the potential function
$\tau_1$ and its derivatives. However, not all of them can be determined by the boundary
relations (\ref{sc4}). In order to determine the remaining data, we have to appeal to 
the central ingredient of Fokas' method, i.e. to derive the relevant {\it global relation}
for the jump function $\rho_{12}(\zeta)$.   

Formula (\ref{uif403}) can be rewritten in the form of the line integral of the conservative vector field,
$$
\rho_{12}(\zeta)=\int_{-\infty < x'< \infty,\, \, \, z'=0}
 e^{-\frac{ih}{2}(\zeta +\frac{1}{\zeta})z'+\frac{h}{2}(\zeta
-\frac{1}{\zeta})x'}[Q_1dz'+{\tilde Q}_1dx']
$$
Assuming that either $\zeta =-it, t > 1$ or $\zeta =it, t<1$, the contour can
be closed in the upper plane $z'\geq 0$.
Therefore, $\rho_{12}(\zeta)$ is zero on these parts of the complex axis $\zeta$,
\beq\label{m7} 
\rho_{12}(\zeta) =0,\quad \zeta = -it,\,\, t >1, \quad\mbox{and}\quad \zeta = it,\,\, 0< t<1,
\eeq
which constitutes the global relation for our problem. 
\begin{figure}[ht]
\begin{center}
\resizebox{3in}{!}{ 
\includegraphics{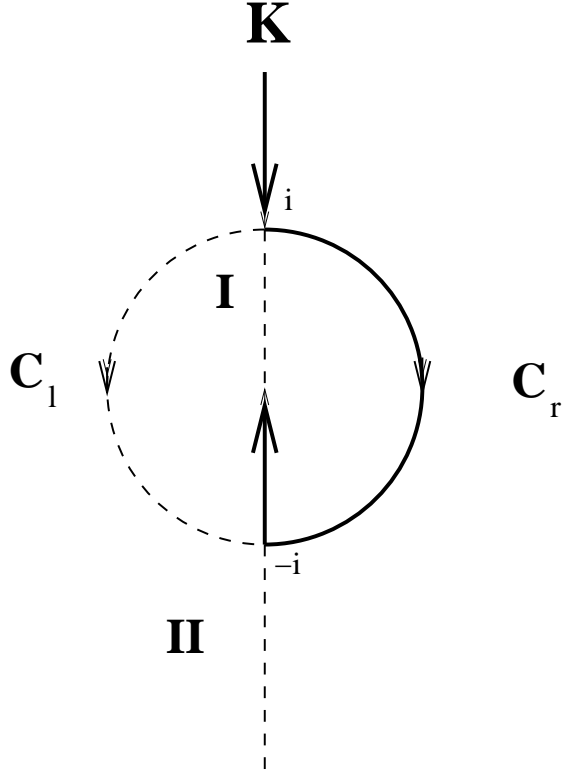}
}
\caption{Contour  of integration $K$ 
of the first scalar Lax pair; sections of the contour where the jump function
is zero are given by the dushed line.}
\label{f3}
\end{center}
\end{figure}
Furthermore, the circular part of the contour has to be analyzed taking into account the radiation
condition. Applying the stationary phase estimate as $R\rightarrow \infty$ ($x=R\cos \theta,\; z= R \sin \theta,
;0\leq \theta \leq \pi$)  to $\tau _1$ (\ref {uif43}) yields two stationary phase points
\beq\label{M4} \zeta_1=\sin \theta - i\cos \theta,\quad \zeta_2= -( \sin \theta - i\cos \theta).
\eeq
They belong respectively to $C_r$ and $C_l$ parts of $K$ (see Fig{\ref{f3}}). These points
provide the following asymptotic estimates 
\beq\label{M5} I_{C_r}\sim \frac{1}{2\pi i} \frac{\rho(\zeta_1)}{\zeta_1}e^{iRh}e^{i\theta}
\sqrt{\frac{2}{Rh}} e^{-i\pi /4} \sqrt{\pi}
\eeq
\beq\label{M6} I_{C_l}\sim \frac{1}{2\pi i} \frac{\rho(\zeta_2)}{\zeta_2}e^{-iRh}e^{i\theta}
\sqrt{\frac{2}{Rh}} e^{-i\pi /4} \sqrt{\pi}
\eeq
The second asymptotic solution (\ref{M6}) does not satisfy the radiation condition (\ref{som});
therefore, in addition to the global relation (\ref{m7}), we have  that
\beq\label{rad}
\rho_{12}(\zeta) =0,\quad \zeta \in C_l.
\eeq

Taking  into account  (\ref{m7}) and (\ref{rad}),  one finally obtains that 
 the jump functions should be defined
on  the ``non zero''  parts of the contour 
 $K$ which are  indicated in  Figure \ref{f3} by solid lines.

\subsection{Analysis of the global relation}

In this section we use the global relation  (\ref{m7}) and the radiation
condition (\ref{rad})
to determine the jump function $\rho_{12}(\zeta)$ in terms of the known
functions $T_{xz}^{0}, \;T_{zz}^{0}$.

Let us rewrite (\ref{uif40333}) changing $x'$ to $x$ and substituting 
$\tilde Q_1$ from (\ref{uif413}):
\beq\label{an44}
\rho_{21}(\zeta)= \int_{-\infty}^\infty e^{\frac{h}{2}\left(\zeta
-\frac{1}{\zeta}\right)x}\left[\left(\frac{i}{h\zeta}\tau _1(0,x)-\frac{i}{h^2}
(\tau_{1x}(0,x)+i\tau_{1z}(0,x))\right)\right]  dx.
\eeq
After integration of $\tau_{1x}$ by parts one obtains
\beq\label{an451}
\rho_{21}(\zeta)= \int_{-\infty}^\infty e^{\frac{h}{2}\left(\zeta
-\frac{1}{\zeta}\right)x}\left[\frac{i}{2h}\left(\zeta+\frac{1}{\zeta}\right)\tau_1(0,x)+
\frac{1}{h^2}\tau_{1z}(0,x)\right]  dx
\eeq
Then using conditions  (\ref{sc4}) at $z=0$, equations (\ref{sc01}), (\ref{sc011}) and
again  integrating by parts,
one finally arrives at the formula,
\beq\label{M1}\rho_{21}(\zeta)=-b(\zeta)\Phi_1(\zeta)-d(\zeta)\Phi_2(\zeta) +F_1(\zeta),
\eeq
where $F_1$ is defined by the given boundary data,
\beq\label{an51}
 F_1=-\frac{i}{4h(\lambda+2\mu)}\left(\zeta + \frac{1}{\zeta}\right)
\int_{-\infty}^\infty e^{\frac{h}{2}\left(\zeta
-\frac{1}{\zeta}\right)x}T_{zz}^{(0)}(0,x)dx
\eeq
$$
-\frac{1}{2l^{2}\mu}\int_{-\infty}^{\infty} e^{\frac{h}{2}\left(\zeta
-\frac{1}{\zeta}\right)x}(T_{xz}^{(0)}(0,x))_xdx
$$
 and 
 $\Phi_1,\; \Phi_2$ are the following  integrals
of the unknown $u$ and $w$:

\beq\label{an561}
\Phi_1(\zeta)=\int_{-\infty}^\infty e^{\frac{h}{2}\left(\zeta
-\frac{1}{\zeta}\right)x}u(0,x)dx, \quad
\Phi_2(\zeta)=\int_{-\infty}^\infty e^{\frac{h}{2}\left(\zeta
-\frac{1}{\zeta}\right)x}w(0,x)dx.
\eeq
The coefficient functions, $b(\zeta)$ and $d(\zeta)$ are given by the  formulas:
\beq\label{an54}
b(\zeta )=\frac{ih^2}{4l^2}\left(\zeta^2-\frac{1}{\zeta^2}\right)
\eeq
\beq\label{an55}
d(\zeta)=\frac{l^2-h^2}{2l^2}+\frac{h^2}{4l^2}\left(\zeta^2+\frac{1}{\zeta^2}\right)
\eeq

In terms of these functions the global relation on the parts I and II of the imaginary $\zeta$ - axis reads
\beq\label{M2}
b(\zeta)\Phi_1(\zeta)+d(\zeta)\Phi_2(\zeta) =F_1(\zeta)
\eeq
Changing $\zeta$ to $-\frac{1}{\zeta}$ and using symmetries yields
\beq\label{M3}
-b(\zeta)\Phi_1(\zeta)+d(\zeta)\Phi_2(\zeta) = F_1\left(-\frac{1}{\zeta}\right)
\eeq
on the parts of  the imaginary axis which are included into non zero $\rho$ sections of $K$. 
Hence the boundary conditions
applied to the first Lax pair produces one equation  to relate the two unknown functions,
i.e.$\Phi_1$ and $ \Phi_2$ on these parts of the oriented contour. 

Equation  (\ref{M2}) also holds  on $C_l$ where $\rho_{12} =0$.
Changing $\zeta$ to $-\frac{1}{\zeta}$ and using symmetries yields
\beq\label{M333}
-b(\zeta)\Phi_1(\zeta)+d(\zeta)\Phi_2(\zeta) =F_1\left(-\frac{1}{\zeta}\right),
\eeq
and hence we obtain an equation (actually the same as (\ref{M3})) relating the two unknown functions on the arc $C_r$ as well.

Repeating  computations for the second Lax pair on the $\tilde\zeta$ complex plane
one obtains that the global relation has similar form as (\ref{M2})

\beq\label{M8}
\delta(\tilde\zeta)\tilde\Phi_1(\tilde\zeta)+\beta(\tilde\zeta)\tilde\Phi_2(\zeta) =F_2(\zeta)
\eeq
where
\beq\label{M9}
\delta(\tilde\zeta)=-\frac{1}{4}\left(\tilde\zeta ^2+\frac{1}{\tilde\zeta^2}\right)
\eeq
\beq\label{M10}
\beta(\tilde\zeta) =\frac{i}{4}\left(\tilde\zeta ^2-\frac{1}{\tilde\zeta^2}\right)
\eeq
\beq\label{M11}
\tilde\Phi_1(\tilde\zeta)=\int_{-\infty}^\infty e^{\frac{l}{2}\left(\tilde\zeta
-\frac{1}{\tilde\zeta}\right)x}u(0,x)dx, \quad
\tilde\Phi_2(\tilde\zeta)=\int_{-\infty}^\infty e^{\frac{l}{2}\left(\tilde\zeta
-\frac{1}{\tilde\zeta}\right)x}w(0,x)dx
\eeq
\beq\label{M12}
 F_2=-\frac{1}{2\mu l^2}
\int_{\infty}^\infty e^{\frac{l}{2}\left(\tilde\zeta
-\frac{1}{\tilde\zeta}\right)x}(T_{zz}^{(0)}(0,x))_xdx
\eeq
$$
+\frac{i}{4l\mu}\left(\tilde\zeta+\frac{1}{\tilde\zeta}\right)\int_{-\infty}^{\infty} e^{\frac{l}{2}\left(\tilde\zeta
-\frac{1}{\tilde\zeta}\right)x}T_{xz}^{(0)}(0,x)dx
$$
Therefore, using the symmetries in the same way as for the first
Lax pair we can obtain another relation between the unknown functions on the non-zero
parts of the contour $\tilde K$ of $\tilde\zeta$ plane. That shows that in order to finish the solution
of the half space problem we only need to transfer both Lax pairs onto the same complex plane.

\subsection{Joint uniformization}

\begin{figure}[ht]
\begin{center}
\resizebox{3in}{!}{ 
\includegraphics{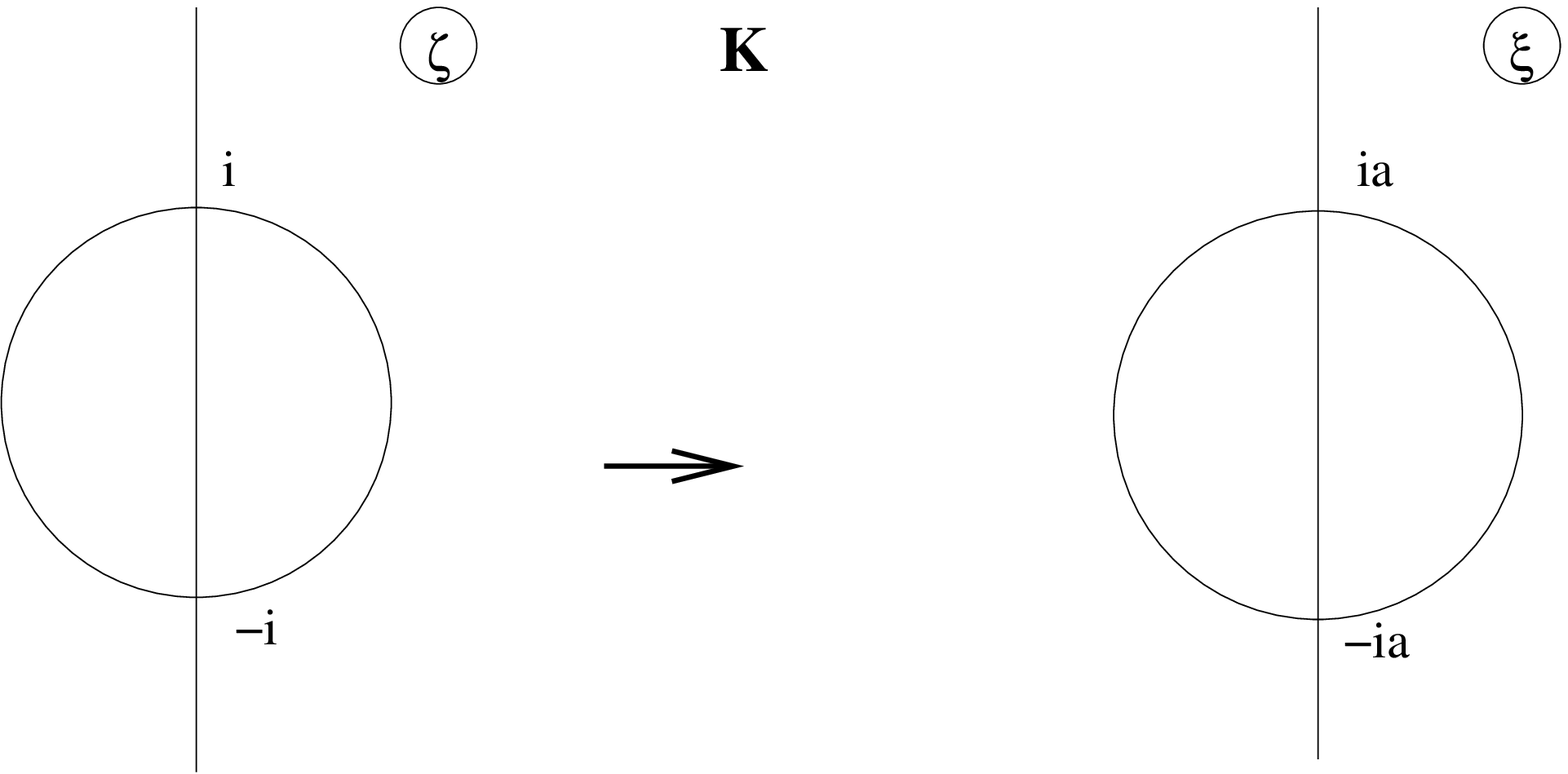}
}
\caption{Transformation of contour  $K$ 
from $\zeta$ to $\xi$ complex plane.}
\label{f4}
\end{center}
\end{figure}

\begin{figure}[ht]
\begin{center}
\resizebox{3in}{!}{ 
\includegraphics{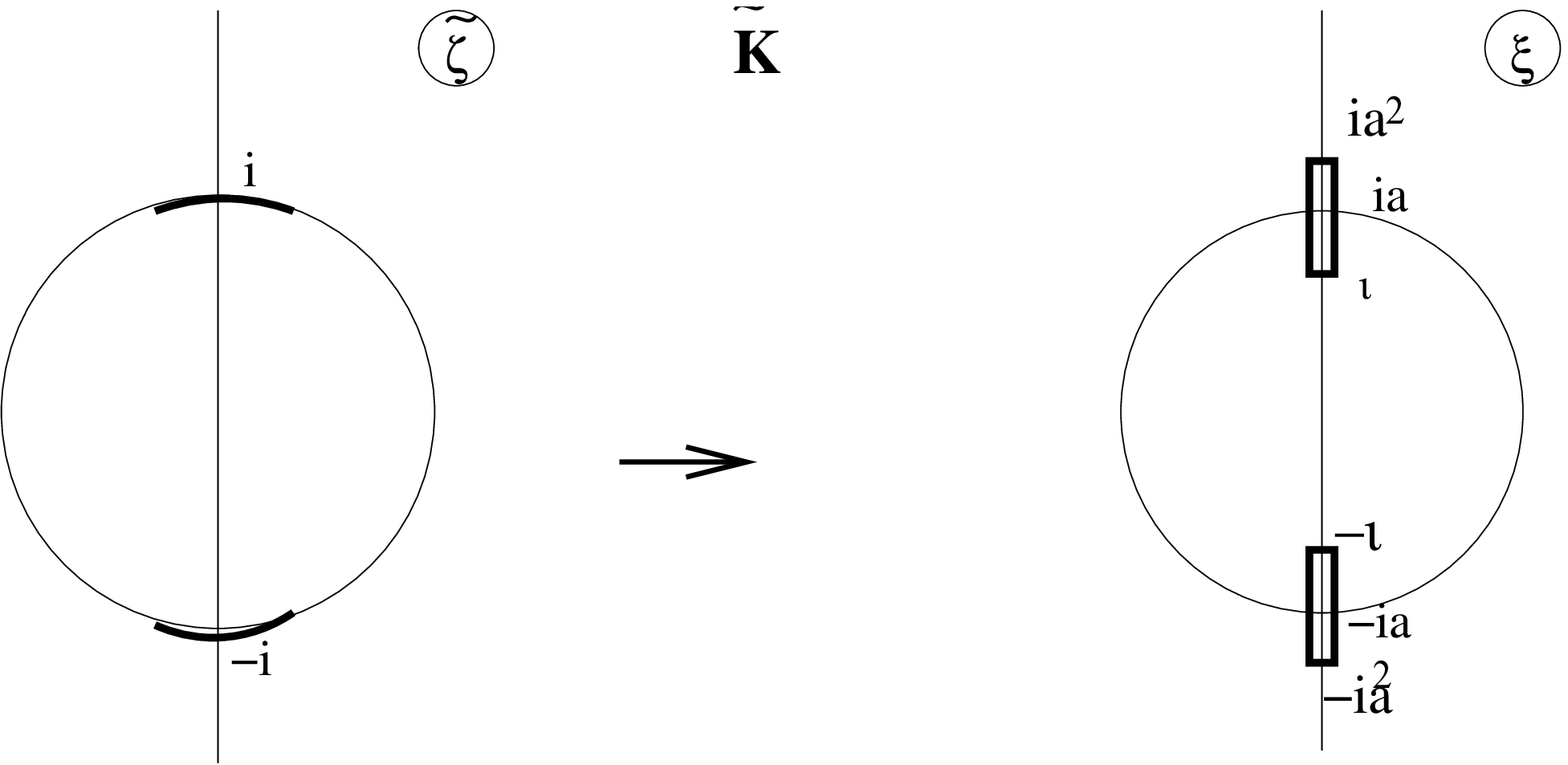}
}
\caption {Transformation of contour  $\tilde K$ 
from $\zeta$ to $\xi$ complex plane.}
\label{f5}
\end{center}
\end{figure}

Let us map the complex planes $\zeta$ and $\tilde\zeta$ to the complex
plane $\xi$ by the following formulae,
\beq\label{M13} \zeta=\frac{\xi}{a}, \quad l\left(\tilde\zeta-\frac{1}{\tilde\zeta}\right)=
h\left(\zeta-\frac{1}{\zeta}\right),
\eeq
where 
\beq\label{M14} a=\frac{l}{h}+\sqrt{\frac{l^2}{h^2}-1}.
\eeq
Transformations of the contours $K$ and $\tilde K$ are presented in Figure{\ref{f4}}
and Figure{\ref{f5}}, respectively. The explicit formula for the map  $\tilde\zeta(\xi)$
is given by the equation,
\beq\label{zetatildexi}
\tilde{\zeta} = \frac{h}{2al}\left(\xi - \frac{a^2}{\xi} +\frac{1}{\xi}
\sqrt{(\xi^2 + 1)(\xi^2 + a^4)}\right),
\eeq
so that
$$
\tilde\zeta + \frac{1}{\tilde\zeta} =  \frac{h}{al\xi}\sqrt{(\xi^2 + 1)(\xi^2 + a^4)}.
$$

Since,
\beq\label{M15} \zeta-\frac{1}{\zeta} =\frac{1}{a}\left(\xi-\frac{a^2}{\xi}\right),\quad
\tilde\zeta-\frac{1}{\tilde\zeta}=\frac{h}{al}\left(\xi-\frac{a^2}{\xi}\right),
\eeq
the both $\Phi_1(\zeta )$ and $\tilde\Phi_1(\tilde\zeta)$ become $\Phi_1(\xi)$ while
the both $\Phi_2(\zeta )$ and $\tilde\Phi_2(\tilde\zeta)$ become $\Phi_2(\xi)$, where
\beq\label{M16}
\Phi_1(\xi)=\int_{-\infty}^\infty e^{\frac{h}{2a}(\xi
-\frac{a}{\xi})x}u(0,x)dx,\quad \Phi_2(\xi)=\int_0^\infty e^{\frac{h}{2a}(\xi
-\frac{a}{\xi})x}w(0,x)dx
\eeq

Taking into account these transformations and changing $b(\zeta),d(\zeta), F_1(\zeta),\beta(\tilde\zeta),
\delta(\tilde\zeta),F_2(\tilde\zeta)$ to $b(\xi),d(\xi), F_1(\x),\beta(\xi),
\delta(\xi),F_2(\xi)$ yields the system of two algebraic equiations for the two unknown
functions $\Phi_1(\xi)$ and $\Phi_2)\xi)$ on all parts of the contour $K(\xi)$. Indeed, we have
that
\beq\label{M17} 
-b(\xi)\Phi_1(\xi)+d(\xi)\Phi_2(\xi) =F_1\left(-\frac{a^2}{\xi}\right),
\eeq
$$
\delta(\xi)\Phi_1(\xi)-\beta(\xi)\Phi_2(\xi) =F_2\left(-\frac{a^2}{\xi}\right)
$$
if $\xi \in [ia^2, i\infty)\cup[ia, ia^2]_+\cup[-ia, -i]_+\cup[-i,i0)\cup C_r$, and
\beq\label{M1711} 
b(\xi)\Phi_1(\xi)+d(\xi)\Phi_2(\xi) =F_1\left(-\frac{a^2}{\xi}\right),
\eeq
$$
\delta(\xi)\Phi_1(\xi)-\beta(\xi)\Phi_2(\xi) =F_2\left(-\frac{a^2}{\xi}\right),
$$
if $\xi \in [i, ia]\cup[-ia, -ia^2]_+$. Here $[\dots ]_{+}$ means  the right side of the
cut $[\dots ]$,  and the functions $b(\xi),d(\xi),\beta(\xi), \delta(\xi)$ are given by the formulae.
\begin{equation}\label{bd}
b(\xi) = \frac{ih^2}{4l^2}\left(\frac{\xi^2}{a^2} - \frac{a^2}{\xi^2}\right),\quad
d(\xi) = \frac{l^2-h^2}{2l^2} + \frac{h^2}{4l^2}\left(\frac{\xi^2}{a^2} + \frac{a^2}{\xi^2}\right),
\end{equation}
\begin{equation}\label{betadelta}
\delta(\xi) = -\frac{h^2}{4l^2}\left[\left(\frac{\xi^2}{a^2} + \frac{a^2}{\xi^2}\right)+\frac{1}{2}\left(a -\frac{1}{a}\right)^2\right],
\quad 
\beta(\xi) = \frac{ih^2}{4a^2l^2}\left(\frac{\xi}{a} - \frac{a}{\xi}\right)\Omega(\xi),
\end{equation}
where
\begin{equation}\label{omega}
\Omega(\xi) = \frac{a}{\xi}\sqrt{(\xi^2+1)(\xi^2+a^4)}.
\end{equation}
It is worth noticing that
\begin{equation}\label{symm}
\Omega(-\xi) = -\Omega(\xi), \quad \Omega\left(\frac{a^2}{\xi}\right) = \Omega(\xi).
\end{equation}

\section{Analysis of the solution. Rayleigh waves}
Summarizing our  derivations, we see that on all parts of the $\xi$ - image of the contour $K$,
the functions $\Phi_{1}(\xi)$ and $\Phi_{2}(\xi)$
can be defined by solving a simple algebraic system. Changing variable $\xi$ back to the
variables $\zeta$ and $\tilde\zeta$, we obtain the jump functions $\rho_{12}(\zeta)$ and 
$\tilde\rho_{12}(\tilde\zeta)$, respectively. This would complete the solution of the half
space problem. Let us look at the solutions of the algebraic systems more carefully.
For example, from (\ref{M17}), it follows, that
\begin{equation}\label{Phisol}
\Phi_1(\xi) = \frac{\beta(\xi)F_{1}\left(-\frac{a^2}{\xi}\right) + d(\xi)F_{2}\left(-\frac{a^2}{\xi}\right)}{D(\xi)}
\quad \mbox{and}\quad   \Phi_2(\xi) = \frac{\delta(\xi)F_{1}\left(-\frac{a^2}{\xi}\right) + b(\xi)F_{2}\left(-\frac{a^2}{\xi}\right)}{D(\xi)},
\end{equation}
where
\begin{equation}\label{det}
D(\xi) = d(\xi)\delta(\xi) -\beta(\xi)b(\xi)
\end{equation}
is the determinant of  system (\ref{M17}). Our task now is to analyze its zeros. 

By a straightforward  calculations, we have that
$$
D(\xi) = -\left(a+\frac{1}{a}\right)^{-1}D_0(\xi), 
$$
\begin{equation}\label{det2}
D_0(\xi) = \frac{1}{4}\left[\left(a-\frac{1}{a}\right)^2 +2\left(\frac{\xi^2}{a^2} +\frac{a^2}{\xi^2}\right)\right]^2
-\frac{1}{a^2}\left(\frac{\xi^2}{a^2} - \frac{a^2}{\xi^2}\right)\left(1 - \frac{a^2}{\xi^2}\right)
\sqrt{(\xi^2 + 1)(\xi^2 + a^4)}.
\end{equation}
Going back to the original spectral parameter,
$$
k = \frac{h}{2}\left(\frac{\xi}{a} + \frac{a}{\xi}\right),
$$
and recalling the definition of the parameter $a$, one can check that
$$
\frac{h}{2}\left(\frac{\xi}{a} - \frac{a}{\xi}\right) = \sqrt{k^2 - h^2}, \quad
\frac{1}{\xi}\sqrt{(\xi^2 + 1)(\xi^2 + a^4)} = \frac{2a}{h}\sqrt{k^2 + l^2 -h^2}.
$$
From this, it is easy to see  that
\begin{equation}\label{det0}
\frac{h^4}{16}D_0(\xi) =  \left(k^2-h^2 +\frac{l^2}{2}\right)^2 - k(k^2 - h^2)\sqrt{k^2 + l^2 -h^2}.
\end{equation}
Introducing the physical quantities (see \cite{Bullen} ),
\begin{equation}\label{phys}
c^2 = \frac{\omega^2}{h^2 - k^2}, \quad \alpha^2 = \frac{\omega^2}{h^2}, \quad \beta^2 = \frac{\omega^2}{l^2},
\end{equation}
we arrive at the final formula for the determinant $D_0$,
\begin{equation}\label{det00}
\frac{c^4h^4}{4\omega^4}D_0(\xi) = \left(2 - \frac{c^2}{\beta^2}\right)^2 - 4 \sqrt{1 - \frac{c^2}{\alpha^2}}\sqrt{1 - \frac{c^2}{\beta^2}},
\end{equation}
which means that
\begin{equation}\label{RR}
D(\xi) = 0 \,\,\,\Longleftrightarrow\,\,\, \left(2 - \frac{c^2}{\beta^2}\right)^2 = 4 \sqrt{1 - \frac{c^2}{\alpha^2}}\sqrt{1 - \frac{c^2}{\beta^2}}.
\end{equation}
Equation in the right hand side of this equivalence relation is the classical equation for the velocity $c$ of the Rayleigh wave - see
e.g., \cite{Bullen}. Hence our main conclusion:

{\it The zeros of the determinant $D(\xi)$ of the linear system (\ref{M17}) representing  the global relation of the half-plane problem coincide with the 
images  $\xi_c$ of the Rayleigh velocity $c$ under the map chain $ c \rightarrow k \rightarrow \xi$.} 

Due to symmetries (\ref{symm}), there are two zeros: 
$$
\xi_c = ia\left(\frac{\alpha}{c} + \sqrt{\frac{\alpha^2}{c^2} -1}\right) = 
i\left(\frac{\alpha}{\beta} + \sqrt{\frac{\alpha^2}{\beta^2} -1}\right)\left(\frac{\alpha}{c} + \sqrt{\frac{\alpha^2}{c^2} -1}\right),\quad\mbox{and}\quad
\frac{a^2}{\xi_c}.
$$
Since $ 0 < c < \beta < \alpha$, the  the roots lie on the intervals $(ia^2, \infty)$ and $(-i, 0)$. This means, that the density $\rho_{12}(\zeta)$ 
has poles on the contour $K$.  Therefore, to ensure that  the boundary value problem under consideration is solvable,
 the corresponding residues of  $\rho_{12}(\zeta)$ must vanish. This   imposes a certain solvability condition on 
 the boundary data, which can be written in the form,
 $$
\left[ \delta(\xi)F_{1}\left(-\frac{a^2}{\xi}\right) + b(\xi)F_{2}\left(-\frac{a^2}{\xi}\right) \right]|_{\xi=\xi_c,\,\,a^2/\xi_c} = 0,
$$
and which can be interpreted  as the orthogonality of the the boundary data to the traces of the Rayleigh waves on the boundary.
This is the mechanism of appearance of the Rayleigh waves  in the solution of the
half-plane within the Riemann-Hilbert approach.  Simultaneously,  we should put  in the right hand sides of the equations 
(\ref{uif43}) and (\ref{uif4333})  the addition  terms of the form 
$$
C_1
e^{\frac{ih}{2a}(\xi_c+a^2/\xi_c)z - 
\frac{h}{2a}(\xi_c-a^2/\xi_c)x} + C_2
e^{\frac{ih}{2a}(\xi_c+a^2/\xi_c)z +
\frac{h}{2a}(\xi_c-a^2/\xi_c)x}
$$
in the expression for potential $\tau_1$, and 
$$
\tilde{C}_1
e^{\frac{ih}{a^2}\Omega(\xi_c)z - 
\frac{h}{2a}(\xi_c-a^2/\xi_c)x} + \tilde{C}_2
e^{\frac{ih}{a^2}\Omega(\xi_c)z +
\frac{h}{2a}(\xi_c-a^2/\xi_c)x}
$$
in the expression for potential $\tau_2$. Those addition terms represent the potentials for the Rayleigh waves propagating 
along the surface $z =0$. 

\


\section*{Acknowledgment}

This work was partially supported by the National Science 
Foundation (NSF) under Grant No. MSS-9313578 and by Grant of London Mathematical Society.

\end{document}